\documentclass[useAMS, usenatbib, amssymb]{mn2e}  
\usepackage{psfig}
\usepackage{graphicx}
\usepackage{epsf}
\usepackage{bm}
\usepackage{lscape}
\title [Third catalogue of CAB]
{A catalog of chromospherically active binary stars (third edition)}

\author[Eker et al.]
       {Z. Eker,${^{1,2}} \thanks{E-mail: eker@comu.edu.tr}$
        N. Filiz Ak $^3$, S. Bilir$^4$, D. Do\u gru$^1$, M. T\" uys\" uz$^1$, E. Soydugan$^1$, \\
\newauthor
         H. Bak\i\c s$^1$, B. U\u gra\c s,$^1$, F. Soydugan$^1$, A. Erdem$^1$, O. Demircan$^1$\\ 
$^1$\c Canakkale Onsekiz Mart University, Faculty of Sciences and Arts, Ulup\i nar Astrophysical Observatory, 17100, \c Canakkale, Turkey\\  
$^2$T\"UB\.ITAK National Observatory, Akdeniz University Campus, 07058 Antalya, Turkey\\
$^3$Erciyes University, Faculty of Sciences and Arts, Department of Astronomy and Space Sciences, Talas Yolu, 38039 Kayseri, Turkey\\
$^4$Istanbul University Science Faculty, Department of Astronomy and Space Sciences, 34119, University-Istanbul, Turkey\\
}      
\date{Accepted 2008 month day.
      Received year month day;
      }
\pagerange{\pageref{firstpage}--\pageref{lastpage}}
\pubyear{2008}

\begin{document}

\maketitle

\label{firstpage}

\begin{abstract}
Chromospherically Active Binaries (CAB) catalogue have been revised and updated. With 203 new identifications, the number of CAB stars is increased to 409. Catalogue is available in electronic format where each system has various number of lines (sub-orders) with a unique order number. Columns contain data of limited number of selected cross references, comments to explain peculiarities and position of the binarity in case it belongs to a multiple system, classical identifications (RS CVn, BY Dra), brightness and colours, photometric and spectroscopic data, description of emission features (Ca II H\&K, $H_{\alpha}$, UV, IR), X-Ray luminosity, radio flux, physical quantities and orbital information, where each basic entry are referenced so users can go original sources. 
\end{abstract}


\section{Introduction}
Chromospherically active binaries (CAB) are the class of binary stars with spectral types later than F characterized by a strong chromosphere, transition region, and coronal activity. Enhanced emission cores of Ca II H\&K lines, and emission filling in the Balmer $H_{\alpha}$ occasionally above the continuum are primary indicators of chromospheric activity and are often accompanied by photometric variability caused by starspots analogous to sunspots. 

Combining RS Canum Venaticorum (RS CVn) stars, first defined by \citet{Hall76} with a preliminary list of 40; later with an improved list of 69 by \citet{Hall81}, and BY Draconis (BY Dra) defined by \citet{BoppFekel77} with 13 binaries, and then including early attempts of collections \citep{NZ84, Eker84, Stras86}, the first version of the CAB catalogue, (CCABS), was published with 168 CAB stars by \citet{Stras88}. In the second version, (CABS), the catalogue was revised and published in the same format with contents of 206 CAB stars by \citet{Stras93}, which is out dated now. 

Especially after {\em Hipparcos} mission \citep{Perryman97}, the number of CAB systems is increased and data quality improved very much. Eker had been compiling catalog data for CAB systems privately even after his contribution \citep{Eker84} to the first version \citep{Stras88}. Containing 284 stars, this unpublished list was the source and starting point of recent studies \citep{Karatas04, Bilir05, Demircan06, Eker06, Eker07, Eker08} at which first observational evidence of dynamical evolution of binary star orbits, most probably driven by orbital angular momentum loss via magnetic braking mechanism \citep{Schatzman59, Kraft67, M68}, was extracted from the galactic space velocities and the kinematical ages of CAB. These studies reminded us once more that collecting catalog data is the first step of doing science. Cataloguing is not only good for directing observers to observe under studied systems and provide basic data, but also is good for reporting most reliable physical parameters and improve statistical significance which could be essential for such specialized studies, and useful for understanding dynamo mechanism, role of magnetic activity in single or binary star evolutions, stellar activity and activity cycles, dynamical evolution etc. Starting from already existing but incomplete list of Eker, which practically contains the two earlier versions of CAB catalogue, we have revised inputs of individual stars in the list and searched new CAB members through the literature and finalized the current version. The catalog data should be considered updated up to December, 2007.

The newly identified CAB stars in the present catalogue were collected from various sources. Basic sources are chromospherically active binary candidate list of \citet{Stras93}, \citet{Stras00}, \citet{Montes01}, \citet{Nordstrom04}, SB9 catalogue of binary stars by \citet{Pourbaix04}, and list of stars claimed to be RS CVn or BY Dra in SIMBAD database which are not in the earlier versions of CAB catalogues. Published articles containing the names of the stars in the tentative list are searched one by one in NASA/ADS services. If binarity and chromospheric activity are confirmed in the published papers, the other input data of the catalogue are searched and gathered. However, we also investigated a limited number of stars, which are binaries, chromospheric activity information was missing or, chromospheric activity was confirmed but binarity was not openly confirmed in the literature. Some of such stars were also included in the present version of the CAB catalogue according to the criteria of selecting CAB stars explained in Section 3.

\section{Contents and differences}
The current version of the catalogue (CABS III) is available only in electronic format. With 203 new identifications, the number of CAB systems is increased to 409. Fig. 1 compares orbital period distribution of current catalogue and previous version \citep{Stras93}. Relative increases of the number of CAB systems at various periods are clear. Fig. 2 displays a similar comparison of maximum visual brightness $(V_{max})$. New identifications sharply increase for $V_{max}>7$ mag. This must be due to increasing accuracies and photometric limit of modern instruments and mostly because of {\em Hipparcos} identifications.
\begin{figure}
\begin{center}
\includegraphics[scale=0.35, angle=0]{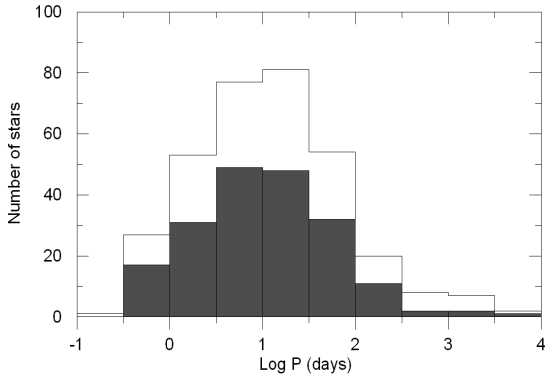}
\caption[] {$\log P$ distributions of CAB in the present catalogue (white bars) and the Strassmeier's et al. (1993) catalogue (black bars).} 
\end{center}
\end{figure}

\begin{figure}
\begin{center}
\includegraphics[scale=0.35, angle=0]{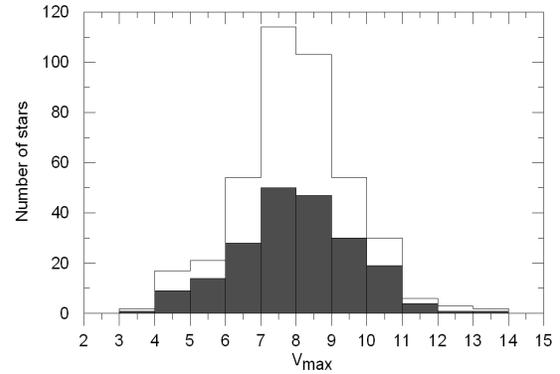}
\caption[] {$V_{max}$ distributions of CAB in the present catalogue (white bars) and the Strassmeier's et al. (1993) catalogue (black bars).}
\end{center}
\end{figure}

The present catalogue contains 1855 rows of data sorted in the order of catalogue number (CABS III) according to increasing right ascension and its sub-orders. More than one line of input data for each star aims to summarize all available specialized studies (columns) for the catalogue stars. Lines of each star are organized such that the most reliable entry is in the first line. Auxiliary information (extra lines) are for researchers or users who like comparing earlier or other entries from different sources. Each entry (column) followed by a column to contain reference numbers. A reference number indicates the original source of the entry. Nevertheless, there are some columns which does not need a reference or there are columns with verbal information followed by a reference number or numbers indicated in a parenthesis. Square parentheses mean the entries are computed by us from the other information available in the present catalogue.

Columns are organized: as order; (CABS III) number; sub-order; star names; maximum brightness in Johnson $V$ magnitude; wave amplitude; photometric, migration and cycle periods; Johnson's colours ($U-B$, $B-V$, $V-R$, $V-I$); spectral types for components; nature of binarity (e.g. SB1, SB2 and SB3), effective temperatures; iron abundance relative to Sun; logarithmic lithium abundances (on a scale with $\log n(H)=12$); projected rotational velocities; Ca II H\&K and H$_{\alpha}$ emissions; luminosity, flux or information of X-Ray, UV, IR and radio; orbital periods and relative orbital period variations; starting phase as a heliocentric Julian date; center of mass velocity; semiamplitude of components (e.g. $K_1$ and $K_2$); eccentricity; longitude of periastron; semi-major axis of apparent orbits; distances from Sun; radii, mass and mass functions; nature of eclipses; inclination of orbits; information of spot modeling; ratio of contributing fluxes between components and wavelength region of this flux ratio; absolute magnitude of the system; magnetic field strength and polarization. The explanations of columns and special entries in the catalogue are given in the electronic form ({\it readme} file in SIMDAB\footnote{Catalogue is only available electronic form at the CDS via anonymous ftp to cdsarc.u-strasbg.fr or via http://cdsweb.u-strasbg.fr/cgi-bin/qcat?J/MNRAS/???/???}). 

\section{Criteria of selecting CAB}
The primary criteria for choosing CAB are in the first paragraph of the introduction. Basically, by definition, there are two necessary conditions; binarity and chromospheric activity, if both are fulfilled for a stellar system, it is sufficient to identify the system as one of the CAB. Despite conditions are that clear, the borderlines of identification could be vague or imprecise.

\subsection{Binarity}  
Among the 409 CAB stars in the catalog, 308 have orbital parameters determined from their radial velocity data, so binarity cannot be doubted. For the rest binarity has been assured either because systems have eclipsing light curves, or dispersions of radial velocity data are confirmed to be much bigger than observational errors of radial velocities. Detection of line doubling and variable radial velocity of components are sufficient to claim such a system to be SB2 even without orbital parameters. For SB1 systems, if there is no evidence of pulsations, variable radial velocity of a single component on the spectrum strongly favors binarity. Statistics of radial velocity data for SB1 systems without orbital parameters are summarized in Table 1. Having maculation wave and photometric periods, those systems are unlikely to be pulsating variables. Data in the table strongly favor those 22 systems are binaries.   

\begin{table*}
\begin{center}
\setlength{\tabcolsep}{2pt}

\caption{Binarity of systems with no orbital parameters.}
{\tiny
\begin{tabular}{clrcccclcccc}
\hline
  CABS III &      &    &  \# of       &             & Typical     & (Min, Max) or  &       &   Wave &            &           \\
   number & Name & HD &  RV          & $\sigma$ RV & error       & [mean] RV    &  Ref.$^{*}$ &  ampl. &$ P_{phot}$ & Comments$^{**}$  \\
      &      &    &  &(km s$^{-1}$)&(km s$^{-1}$)&(km s$^{-1}$) &       &  (mag) &  (days)    &           \\

\hline
           3 & BD+45 4408 &         38 &         10 &       1.0   &       0.4  & (-0.13, 2.76) &          1 &            &            & i, ii \\
          45 &  BD+06 411 &      16884 &          3 &       27.6 &          2.0 & (-25.7, 23.9) &          2 &       0.06 &            & iii \\
          50 &     EP Eri &      17925 &         22 &       2.0 &        1.6 & (14.9, 21.7) &          3 &       0.05 &       6.85 &        \\
          64 &   V969 Tau &      23386 &          5 &        4.6 &          2.0 &      [8.6] &          4 &      0.045 &      2.377 &        \\
          99 &  BD+52 982 &      37216 &          4 &        0.3 &        0.2 &       [11] &        4,6 &            &            &         \\
         103 &     TZ Col &      39576 &          4 &        7.8 &        3.5 &  (12.9, 30.4) &          5 &       0.06 &       2.70 & i, ii \\
         104 & CD-57 1292 &      39937 &          4 &       14.2 &        7.1 &      [1.0] &          4 &      0.005 &            &   ii \\
         143 & BD+28 1600 &      71028 &            &            &            &            &            &       0.02 &       8.35 & iv \\
         152 &     CY Pyx &      78644 &          2 &            &            & (-24.4, 34.3) &          2 &       0.09 &       0.84 & iii \\
         157 &   V414 Hya &      81032 &            &            &            &            &            &      0.288 &     18.802 & v \\
         160 & BD+11 2052B&            &          2 &            &            & (27.1, 47.2) &          2 &            &            & iii \\
         172 &     EQ Leo &            &          2 &            &            &   (8.9, 18.3) &          2 &       0.14 &      33.29 & iii \\
         173 &     LX UMa &      88638 &          2 &        2.5 &        1.7 &     [41.0] &            &      0.052 &      4.935 &            \\
         201 &   V838 Cen &     102077 &          2 &        0.4 &        0.3 &     [16.8] &          4 &       0.04 &      1.848 & vi \\
         214 & BD+07 2588 &     112099 &          2 &            &            & (-26.6, -22.6) &          2 &            &            & iii \\
         218 &     CD CVn &            &          2 &            &            & (-51.8, -41.7) &          2 &      0.125 &      38-42 & iii \\
         263 & 1E1548.7+1125 &         &          4 &      20.6 &          5.0 & (-70.99, -20.61) &          7 &            &            & i, ii \\
         265 &   V383 Ser &     142680 &          3 &      29.2 &        6.2 & (-111.6, -53.2) &          2 &      0.032 &      33.52 & iii \\
        282 &     GI Dra &     150202 &          2 &            &            &  (3.6, 12.3) &          2 &            &            & iii \\
        318 & 2E1848.1+3305 &          &          8 &      27.9 &          5.0 & (-40.5, 38.9) &          8 &            &            & ii \\
        321 &     PZ Tel &     174429 &          9 &        5.2 &        1.7 &     [-5.8] &          4 &    0.1-0.2 &    0.94583 &            \\
        346 &  V4429 Sgr &     190642 &          3 &       11.5 &        1.4 & (-25.8, -2.8) &          2 &            &            & iii \\
\hline
\end{tabular} 
}
\end{center}
$^{*}$(1) \citet{Tokovinin02}, (2) \citet{Stras00}, (3) \citet{Beavers86}, (4) \citet{Nordstrom04}, (5) \citet{Buckley87}, (6) \citet{Stras00}, (7) \citet{Silva85}, (8) \citet{Takalo88}

$^{**}$ (i) CCABS, (ii) CABS, (iii) \citet{Stras00} notted ``newly discovered SB1'', (iv) Variable RV \citep{Heard56, Stras94}, (v) Binarity claimed by \citet{Pandey05}, (vi) Visual binary, $P\sim 1$ yr \citep[Hipparcos,][]{Perryman97}
 \end{table*}

The shortest orbital period is 0.309 days (DV Psc) and the longest orbital period is 9166 days (HD 3443). Towards the lower limit of orbital periods, a special care was given not to include contact binaries of W UMa type since they have been recognized as a distinct class, even though significant fraction of W UMa binaries are known to have chromospheric activity \citep{Rucinski86}. At first, we have intended to exclude semi-detached systems too but later we have decided to keep them in because there are un-ignorable numbers of such systems in the earlier versions of the catalogue, even in the original list of \citet{Hall76}. Nevertheless, unlike the earlier versions, in this version we have preferred to indicate semi-detached status or any other peculiarity exist such as membership to moving groups, associations or clusters, mass transfer, multiplicity etc. Such information are given in the file {\it comments.dat} in the electronic form, where also the position of the CAB is described within the multiple system which could also be a visual binary or a visual multiple. 

Following the pioneering works by \citet{Wilson63}, \citet{Kraft67} and \citet{Skumanich72}, which led to the hypothesis that rotation plays a crucial role in generating stellar magnetic activity, binarity is recognized as main reason for CAB to have higher chromospheric activity than normal single stars because high rotation rate of binary components are maintained due to tidal interactions between the binary components. Although tidal interactions become weaker at longer orbital periods, it may appear very long periods (e.g. longer than few hundreds of days) not important for the chromospheric activity. Since there is no definite upper limit especially for definition of CAB, we decided not to discard very long period systems too as long as orbital parameters are determined from radial velocity variations. 

Long period spectroscopic binaries, if not too distant, could also be observed as visual binaries; e.g. Capella, OR Del, 4 UMi, HD 132756. If no radial velocity variation was detected and orbital parameters were not confirmed spectroscopically, visual binaries and proper motion binaries are excluded from the list. Nevertheless, a CAB could be within a multiple system, which itself could be a spectroscopic, visual or proper motion multiple or all possible combinations.

\subsection{Chromospheric activity}
Various techniques had been described in literature for determining Ca II H\&K emissions. Most popular ones are Mount Wilson $S$ index \citep{Wilson68,Duncan91}, and $R_{CaII}$ index \citep{Linsky79, Stras00} and various empirical indexes defined by various authors \citep{Hall96, Twarog95, Montes95, Wright04, Cincunegui07, Hall07}. In the first version \citep{Stras88} Wilson's criteria, where various activity levels indicated from $I_{K}$=1 to 5 according to eye-estimated scale, was accepted and systems with $I_{K}>2$ were chosen into the catalogue. In the second version \citep{Stras88}, absolute Ca II H\&K emission-line surface fluxes defined as $R_{Ca II}=(F^{'}(H_{1})+F^{'}(K_{1}))/\sigma T^{4}_{eff}$
were used and systems with $R_{CaII}>2-4\times10^{-5}$ are included in the catalogue. 

For this version of CABS III we have preferred to be more flexible and include all systems if authors confirm chromospheric activity not only using any of the methods for detecting Ca II H\&K emissions but also according to $H_{\alpha}$, X-Ray luminosity or star spots. Information about Ca II H\&K emission is dedicated in the column named {\it Ca\_em} (in electronic form) where only simple descriptions are given because there is no standard way of expressing flux values.

Among the 409 systems in this version, there exist Ca II H\&K information for 387. There are only these 22 systems with no information about Ca II H\&K emission. Activity properties of 22 systems are summarized in Table 2. All systems in the catalogue are given a defining reference under the column {\it references for class} (see {\it readme} file in the electronic version). 

\begin{table}
\setlength{\tabcolsep}{2pt}
\begin{center}
\caption{Activity of systems with no information about Ca II H\&K emission.}
{\scriptsize
\begin{tabular}{cclcccccc}
\hline
       CABS III &            & H$\alpha$ & Phot. &      X-ray &            &       \\
         number &       Name & activity$^{*}$ &   wave &   activity &      Class &    Defined$^{**}$ \\
\hline
             40 &   V405 And  & se & $\surd$ & $\surd$ &     BY Dra &          1  \\
             60 &     IX Per  &    & $\surd$ & $\surd$ &     RS CVn &          2  \\
             68 &     DF Cam  &  e & $\surd$ & $\surd$ &     RS CVn &          3  \\
             90 &     EZ Eri  & cf & $\surd$ & $\surd$ &            &          4  \\
             93 &     VV Lep  &    & $\surd$ &         &     RS CVn &          5  \\
             94 &  V1236 Tau  &  e &         &         &            &          6  \\
            120 &     VV Mon  &  e &         & $\surd$ &     RS CVn &         16  \\
            127 &   V340 Gem  & cf & $\surd$ & $\surd$ &     RS CVn &          3  \\
            144 &     FF Cnc  &    & $\surd$ &         &            &          7  \\
            179 & BD+38 2140  &  e & $\surd$ & $\surd$ &            &          3  \\
            189 &     TV Crt  &  e & $\surd$ & $\surd$ &     BY Dra &          9  \\
            193 &     VV Crt  &    & $\surd$ & $\surd$ &            &          4  \\
            194 &   V858 Cen  & cf & $\surd$ & $\surd$ &     RS CVn &         10  \\
            244 &     NZ Vir  &    & $\surd$ & $\surd$ &            &         11  \\
            253 &     GU Boo  & se & $\surd$ & $\surd$ &     BY Dra &         12  \\
            269 &GSC 2038-293 &    & $\surd$ & $\surd$ &     RS CVn &         13  \\
            287 &  V1034 Her  &    & $\surd$ &         &     RS CVn &         14  \\
            338 & BD+11 3873  & cf &         & $\surd$ &            &          3  \\
            350 & BD+44 3402  &  e &         & $\surd$ &            &         15  \\
            380 &   V376 Cep  &    & $\surd$ & $\surd$ &     RS CVn &      17,18  \\
            381 &     OT Peg  & se & $\surd$ & $\surd$ &     BY Dra &          3  \\
            386 & BD+33 4462  &  e & $\surd$ & $\surd$ &     RS CVn &          3  \\
\hline
\end{tabular}
}
\end{center}

$^{*}$ se: strong emission, e: emission, cf: core filling

$^{**}$(1) \citet{Chevalier97}, (2) \citet{Drake89}, (3) \citet{Frasca06}, (4) \citet{Cutispoto96}, (5) \citet{Griffin06}, (6) \citet{Bayless06}, (7) \citet{Robb96}, (8) \citet{Frederic96}, (9) \citet{Fekel93}, (10) \citet{Cutispoto98}, (11) \citet{Robb97}, (12) \citet{Lopez05}, (13) \citet{Bernhard06}, (14) \citet{Kaiser02}, (15) \citet{Osten98}, (16) \citet{Hall76}, (17) \citet{Drake92}, (18) \citet{Makarov03}

\end{table}
{\tiny

}

\section{Completeness}
The catalogues, especially such as CAB, W UMa, FK Com, RR Lyr, cepheids, cataclysmic variables etc., are never complete. This is because not only there are always new identifications and even if there is no new identifications there are new  accurate observations and newer studies, but also because there would be some missing information (columns), at which some scientists would wish to have depending upon their special interests. Here we were careful to present at least very basic observational and first order evaluation of observations such as orbital parameters and physical quantities of the systems. Missing second order data, e.g. space distributions and velocities, evolutionary, isochronal or kinematical ages, we think, must be left out because such information can be derived from the present data later according to the personal style of researchers who would prefer to re-derive them. Some second order data may require specialization, e.g. X-Ray, UV, radio etc.

Fig. 3 compares period histogram of northern ($\delta>0^{\circ}$) and southern ($\delta<0^{\circ}$) stars, where understudied nature of southern are displayed clearly. We encourage observers to give extra care to study more southern systems in order to improve homogeneity in the sky distribution, which is crucial for studies such as space distributions, luminosity functions, space velocities, kinematics ages of CAB in the solar neighbourhood   

\begin{figure}
\begin{center}
\includegraphics[scale=0.35, angle=0]{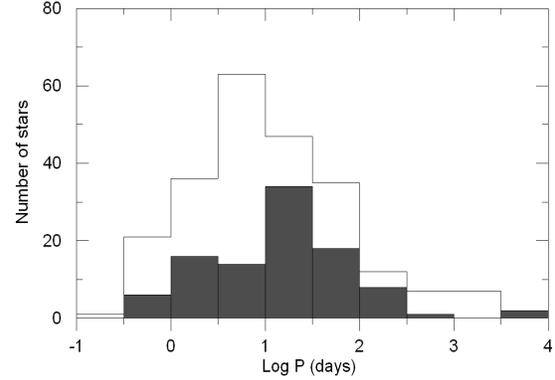}
\caption[] {Comparing period histogram of northern ($\delta>0^{\circ}$, white bars) and southern ($\delta<0^{\circ}$,  black bars) CAB systems in the present catalogue.}
\end{center}
\end{figure}

\section{Acknowledgements}
This work has been supported by the Scientific and Technological Research Council (T\"UB\.ITAK) 104T508. We thank Edwin Budding, Didem T\" uys\" uz, Mukadder I\u gdi \c Sen, Emine Koparan for their helps and varoius contributions. This research has made use of the SIMBAD database, operated at CDS, Strasbourg, France and NASA's Astrophysics Data System.

\end {document}